***Updated version.***

# Tunable dislocations overcome mechano-functional tradeoff in perovskite oxides

Jiawen Zhang[1], Wenjun Lu[1]*, Xufei Fang[2]*

[1]Department of Mechanical and Energy Engineering, Southern University of Science and Technology, Shenzhen, 518055, China

[2]Institute for Applied Materials, Karlsruhe Institute of Technology, Karlsruhe 76131, Germany

*Corresponding authors: luwj@sustech.edu.cn (W. Lu); xufei.fang@kit.edu (X. Fang).

**Abstract:**

Recent advancements in dislocation engineering are reshaping the traditional view towards ceramics being brittle. Here, we use $KTaO_3$ (KTO), a perovskite oxide that is newly discovered with room-temperature bulk plasticity, and demonstrate that the seeded dislocations can effectively tune both mechanical and functional properties. We uncover a novel brittle-ductile-brittle (BDB) transition: low dislocation densities lead to brittle failure, intermediate densities (~$10^{14}$ $m^{-2}$) enable superior ductility with strains over 20%, and high dislocation densities (~$10^{15}$ $m^{-2}$) induce again brittle fracture. This dislocation density-dependent non-monotonic mechanical response challenges the traditional behavior of ceramics and offers new design opportunities. Furthermore, dislocation densities can monotonically decrease thermal conductivity, revealing a tradeoff between mechanical strength and functionality. The findings reveal a critical threshold of dislocation density in optimizing the performance of functional oxides, and provide a new framework for using dislocations to design advanced materials where mechanical durability and enhanced functionality are intertwined.

***Updated version.***

## I. Introduction

Ceramics are typically categorized as brittle materials, exhibiting negligible plastic deformation under ambient conditions. This perception, long embedded in both scientific understanding and industrial design, stems from the strong and directional ionic/covalent bonding that causes high lattice friction stress and insufficient operative slip systems at room temperature [1]. The direct consequence is a lack of easy dislocation (1D crystalline defect) glide in ceramics, as compared to metallic materials [1]. Up to date, the potential of ceramics has been shackled to passive structural applications, where strength and hardness are prioritized over ductility and toughness [2] despite the fact that the latter has been pursed continuously [3, 4]. Meanwhile, tuning the functional properties of ceramics have been limited to 0D point defect engineering via defect chemistry [5, 6] or 2D boundary engineering through interface design [7] or texturing [8, 9]. Dislocation as a fundamental defect type has unfortunately long been cast out for ceramics engineering. However, recent breakthroughs in dislocation engineering in oxides [10], particularly *dislocation seeding* [11] involving direct mechanical imprinting into ceramic oxides, have begun to fundamentally challenge this conventional view and open new frontiers in the mechanical and functional design of oxides [12, 13], with a potential to be extended to more than dozens of ductile ceramics that exhibit room-temperature bulk plasticity mediated by dislocations [14].

Perovskite oxides, because of their rich electronic properties and structural tunability, have become archetypal platforms for exploring dislocation-based novel properties [12, 15-18]. Among them, $SrTiO_3$ (STO) and $KTaO_3$ (KTO), as prototypical functional oxides, have attracted particular attention for their capacity to host dense dislocation networks by bulk compression at room temperature [19-23]. These materials not only exhibit tunable electronic and electromechanical responses to dislocations, but also serve as model systems to investigate room-temperature bulk plasticity in crystalline oxides from a fundamental perspective. For instance, the current authors have recently demonstrated that STO can be mechanically conditioned to exhibit large room-temperature plasticity with more than 30% plastic strain via a strategy with controlled dislocation seeding [11] to overcome dislocation nucleation, a common bottleneck for achieving plasticity in ceramics. By incrementally introducing dislocations using cyclic scratching, we have achieved dislocation densities over five orders of magnitude ranging from ~$10^{10}$ to $10^{14}$ $m^{-2}$ in single-crystal STO, without triggering grain refinement or phase instability at room temperature [24]. This enables systematic investigations of the relationship between dislocation density and mechanical response in a structurally cleaner system [24].

A key finding from this line of research was the emergence of the non-monotonic mechanical behavior: the yield strength of STO initially decreased as dislocations were introduced, attributed to enhanced mobility and dislocation glide, but then the yield strength increased with higher dislocation densities due to forest hardening and dislocation-dislocation interactions [24-26]. This "V-shape" behavior in strength (**Figure 1a**), reminiscent of dislocation nucleation, multiplication, and work-hardening phenomena in metals [27], marked a conceptual breakthrough in ceramics as it challenged the brittle paradigm of oxide ceramics and provided a new and fundamental pathway toward tailoring ductility through tunable dislocation densities [10, 24]. However, the state-of-the-art dislocation density plateaued in STO around $10^{14}$ $m^{-2}$, beyond which further densification was hindered by brittle fracture, limiting access to the full mechanical landscape.

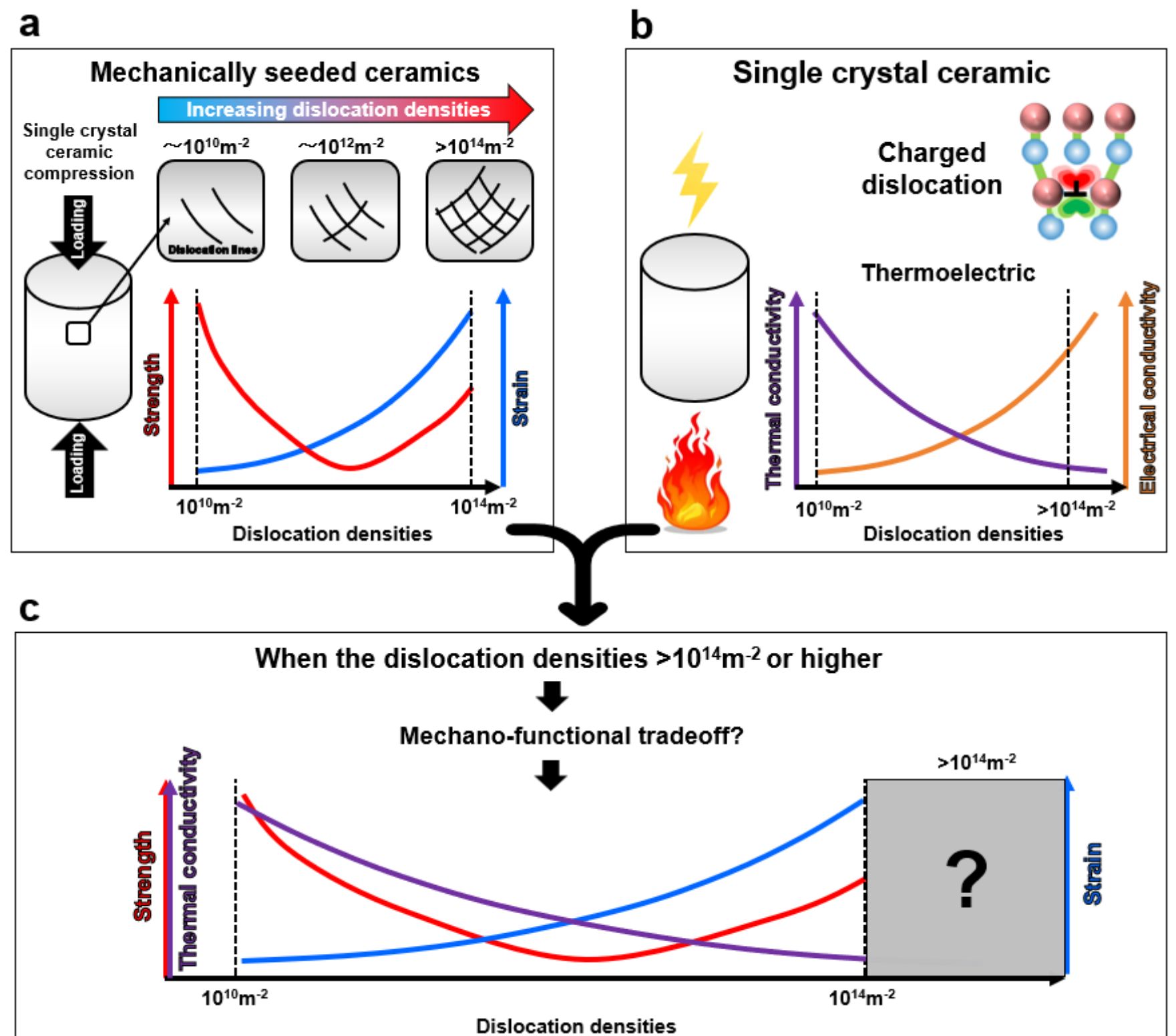

**Figure 1.** Mechanically seeded dislocations in brittle functional single-crystal ceramic materials. (a) Strength and strain as function of the tunable dislocation densities in functional ceramics (with a maximum density of ~$10^{14}$ $m^{-2}$ as state of the art [24]); (b) Electrical and thermal conductivity as function of dislocation density; (c) Beyond the limit: what changes in mechanical properties would occur if the dislocation density surpasses the

***Updated version.***

limit of ~$10^{14}$ m$^{-2}$ ? Note the mechano-functional properties tradeoff as function of dislocation density.

Equally important, the mechanics-enabled dislocation engineering approach has catalyzed the assessment of dislocation-tuned functional properties [28]. Driven by the new research horizon of *dislocation technology* in functional oxides [10, 13, 28, 29], one rising research trend has been centered on pushing the limit of dislocation density in ceramics to maximize the physical properties, for instance, boosting electromechanical properties in ferroelectrics [12], enhancing the electrocatalytic performance by way of dislocation-modified hydrogen adsorption free energy [30], increasing electrical conductivity by overlapping the space charge zones around charged dislocations (cores) [31, 32], and decreasing thermal conductivity due to effective phonon scattering [33], with the latter two combined to boost thermoelectric figure of merit for energy harvesting [34, 35]. A consensus appears to be that the higher the dislocation density, the better the outcome of these functional properties, reflected by a monotonic correlation between them, as schematically represented for electrical and thermal conductivity in **Figure 1b**.

Yet up to date, these two pieces of the puzzle concerning the mechanical and functional aspects have not been put together due to these two much divided research fields. When fusing these two aspects (**Figure 1c**), namely, non-monotonic change in strength but monotonic variation in physical properties as a function of increasing dislocation density, clearly a tradeoff needs to be sought. This tradeoff will be crucial for the integration of mechanical integrity and functional performance for designing next-generation dislocation-based devices.

To establish a model case, here we used KTO as a new prototypical perovskite oxide (**Figure S1**) alongside STO to assess the dislocation-tuned functional properties in combination with the mechanical properties. KTO has recently been discovered to exhibit large room-temperature bulk plasticity by the current authors [21] in parallel with the Greven's group [22]. A new set of bulk compression test for KTO is also showcased in **Figure S2**. Analogous to STO, KTO is cubic at room temperature and is one of the most widely studied perovskite oxides that have received extensive research in recent years for its versatile structural and functional versatility [21-23, 36, 37]. Nevertheless, compared to STO, distinct structural and electronic properties exist in KTO. The latter possesses a larger lattice parameter, enhanced chemical stability, and a more covalent Ta-O bonding character, all of which influence its dislocation behavior differently than STO [22, 23]. KTO is also used to address the generality of the dislocation seeding approach recently validated in STO [11].

**Results and Analyses**

As illustrated in **Figure 2a**, the high tunability of dislocation densities in KTO enabled a remarkably extended regime beyond ~$10^{15}$ $m^{-2}$ (at least one order of magnitude higher than that in single-crystal STO [24]). This unprecedentedly high value reveals additional features to the non-monotonic mechanical response: as the dislocation density increases, the material goes through a brittle-ductile-brittle (BDB) transition, as evidenced by the micropillar compression. With a low dislocation density (**Figure 2a1-d1**), brittle fracture ensues right after the elastic deformation, as evidenced by the abrupt failure indicated in the stress-strain curve and the completely crushed pillar after deformation.

As the dislocation density increases, a notable shift occurs: the stress-strain curves exhibit smooth yield and clear work hardening behavior, with total strains exceeding ~15% (**Figure 2a2-d2 and Figure S3a**) and ~20% (**Figure 2a3-d3**), indicative of stable dislocation glide and strain accommodation. When the dislocation density exceeds a critical threshold (~$10^{15}$/$m^2$), brittle fracture re-emerged (**Figure 2a3-d3, a4-d4, and Figure S3b**). This dislocation density dependent BDB phenomenon has not been reported before for oxide ceramics. The fracture/yield strength exhibit a similar but more pronounced non-monotonic change as a function of the dislocation density. Both mechanical responses (strength and strain) suggest that excessively high dislocation density destabilizes plastic flow and increase the chance for cracks to occur. It is worth emphasizing that even when the dislocation density in KTO reaches ~$10^{15}$ $m^{-2}$, the crystal structure remains single-crystalline, without undergoing polycrystallization or amorphization (**Figure S4**). This remarkable structural stability highlights the intrinsic resistance of KTO against defect-induced structural degradation, underscoring a fundamental distinction between complex oxides and conventional metallic systems. In addition, as shown in **Figure S5**, increasing the pillar size up to approximately 3 μm leads to noticeable changes in the yield strength of the samples, while the overall deformation mechanisms and failure modes remain essentially unaffected. This observation suggests that, although larger dimensions can influence the onset of plasticity [38, 39], the fundamental deformation behavior of the material is size-independent within this range. The underlying mechanisms will be discussed later based on the dislocation pileup and work hardening.

***Updated version.***

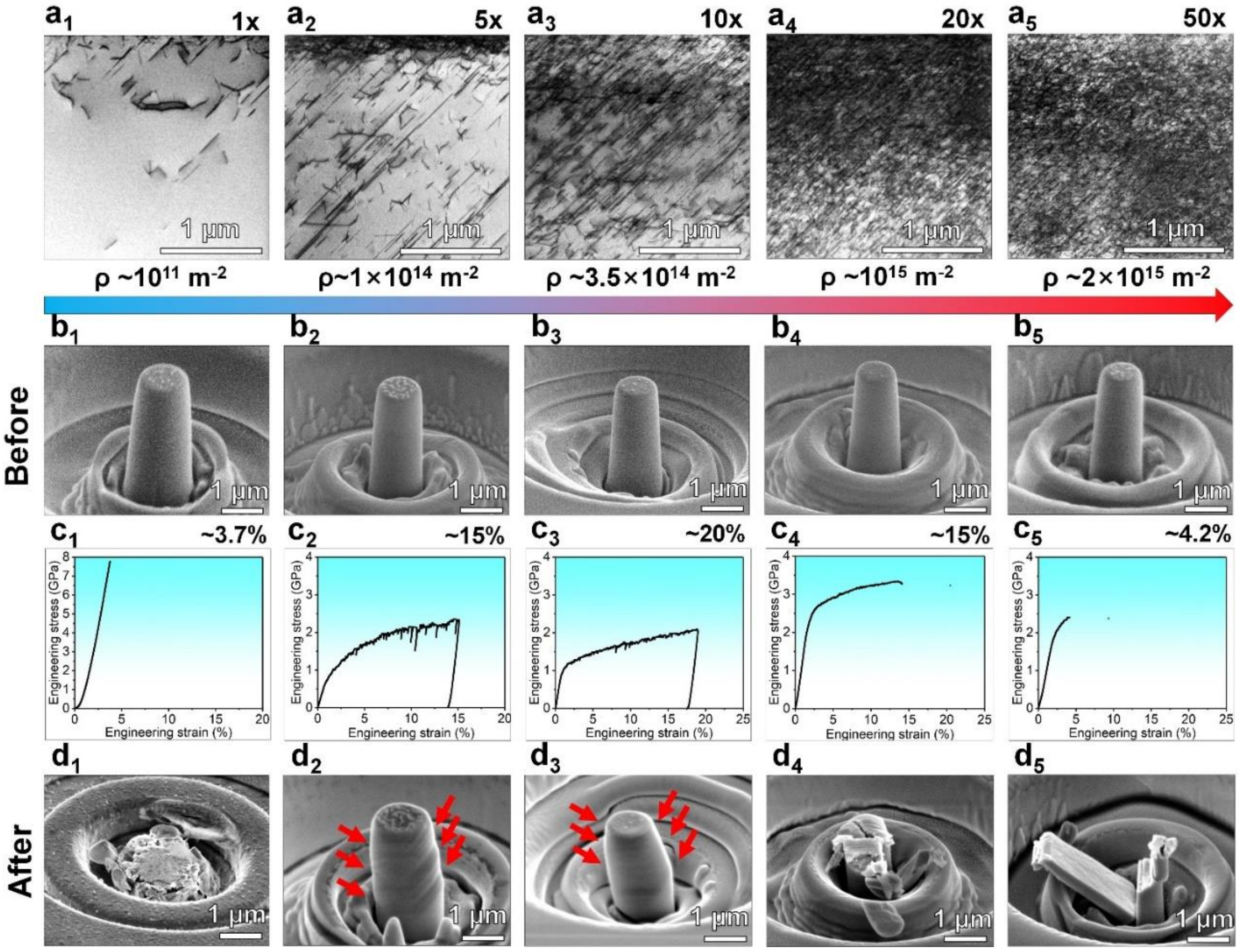

**Figure 2.** Dislocation density tuning and micropillar compression tests in dislocation-seeded samples. Dislocation density tuning is achieved by using a Brinell indenter for scratching the sample surface with different passes (i.e., via cyclic scratching, details in **Materials & Methods**). (a1–a5) Annular bright filed (ABF)-scanning transmission electron microscopy (STEM) images showing a progressive increase in dislocation density beneath the scratch area with increasing number of scratch passes (1× to 50×, note 1× means 1 pass), ranging from $\sim 10^{11}$ to $\sim 10^{15}$ m$^{-2}$. (b1–b5) Scanning electron microscopy (SEM) images of micropillars with varying seeded dislocation densities before *ex situ* compression, which eliminates the electron beam effect. (c1–c5) Engineering stress–strain curves of the deformed micropillars. (d1–d5) SEM images of micropillars after compression, with red arrows in d2 and d3 indicating the slip traces for dislocation activity, while those with much lower dislocation density in d1 shows completely brittle shatter, and these with ultra-dense dislocations in d4 and d5 display again crack and splitting in a brittle manner.

To directly visualize the deformation processes and shed light on the mechanisms governing the BDB transition in KTO, we further conducted *in situ* nanopillar compression experiments in an aberration-corrected STEM. Representative engineering stress-strain curves (**Figures 3a–c**) and corresponding snapshots confirm the influence of dislocation density on the various deformation responses. The pristine sample (1× scratched, **Figure 3a**) exhibits classic brittle fracture at a total strain of ~9%, with no evidence of plastic yield or dislocation activity prior to the catastrophic brittle failure (**Figure 3a5**). In contrast, the 10× scratched sample (**Figure 3b**) demonstrates remarkable plasticity, sustaining over 30% compressive strain without cracking. The snapshots from the *in situ* tests (**Figures 3b1–b4**) capture the continuous formation and propagation of dense slip bands, signifying profuse dislocation multiplication and strain delocalization. This *ductile* regime, enabled by an intermediate dislocation density (~$10^{14}$ $m^{-2}$), aligns with the observations in micropillar tests. Continued increase in the dislocation density (20× scratched sample, **Figure 3c**) leads to an early onset of fracture at a total strain of ~18%. Although dislocation slip is still visible (**Figures 3c1–c4**), the deformation becomes less homogeneous, with localized shear preceding failure. These findings confirm the non-monotonic mechanical response as a function of dislocation density: an initial enhancement of plasticity via mobile dislocation networks is followed by a re-emergence of brittleness due to dislocation pileup and dynamic jamming.

To further address the size effect in nano-/micropillar compression, as most widely discussed in metals [38, 39], we perform systematic size-dependent tests (**Figures 3d–f and Figures S6-8**) to confirm that this BDB transition persists across varying nanopillar diameters. The results suggest that this dislocation density dependent BDB behavior is an intrinsic feature induced by dislocations and not limited by dimensional constraints or the pillar size. These real-time *in situ* observations in nanopillars (**Figure 3**) align with the *ex situ* micropillar compression tests (**Figure 2**) and confirm the role of dislocation-mediated mechanisms in tuning plasticity in KTO, highlighting the critical density window in which optimal mechanical properties can be achieved.

***Updated version.***

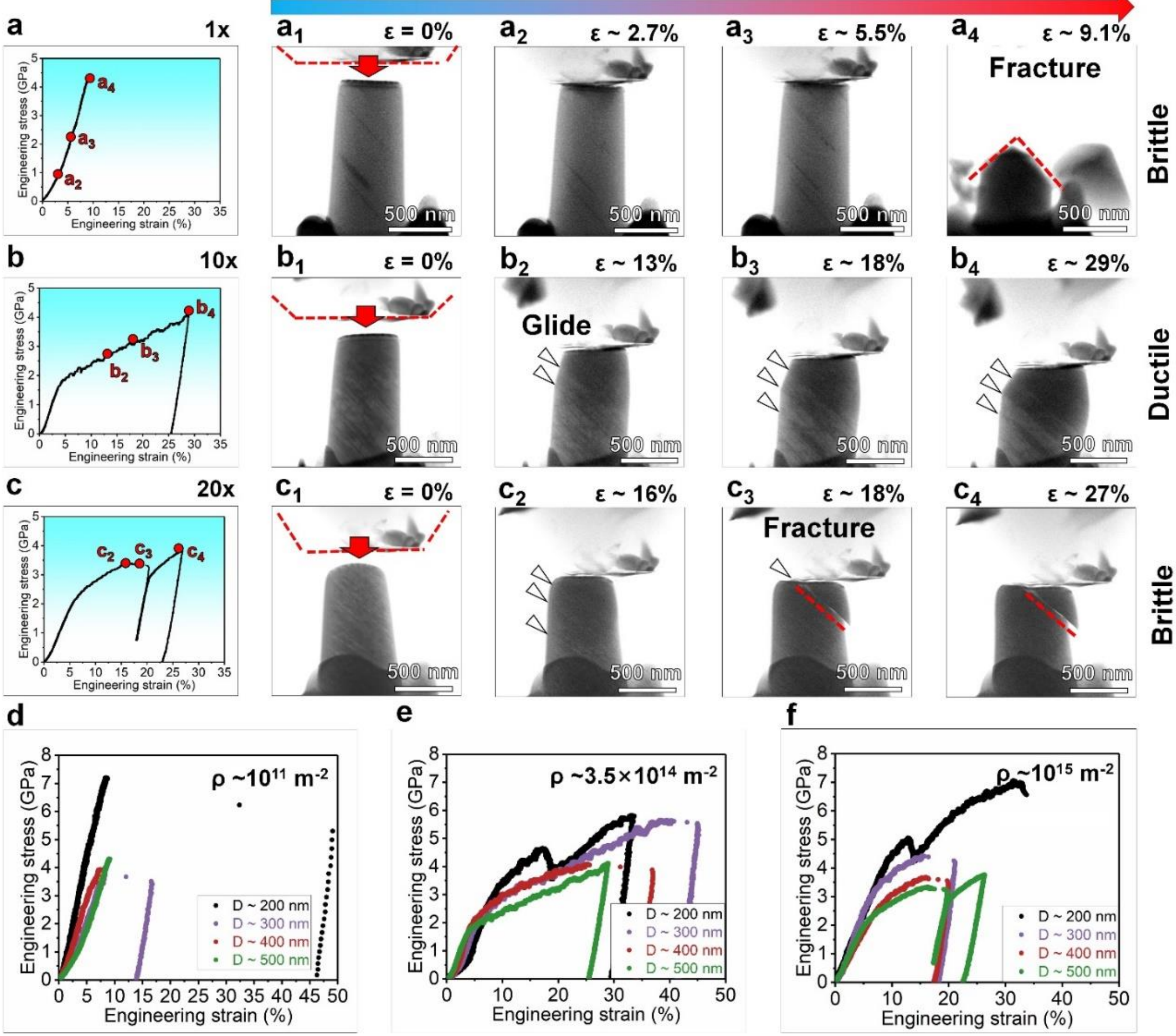

**Figure 3.** *In situ* nanopillar compression in ABF-STEM imaging mode: (a) Engineering stress-strain curve for sample with 1× scratch, (a1–a4) corresponding image snapshots from the *in situ* test, where fracture occurs at a maximum strain of ~9%. (b) Stress-strain curve of the nanopillars in the 10× scratched sample, (b1–b4) corresponding image snapshots visualizing numerous slip bands (indicated by the white arrows), with the nanopillar achieving up to ~30% strain without fracture. (c) Stress–strain curve of the 20× scratched sample, (c1–c4) corresponding image snapshots reveal fracture occurs at a maximum strain of ~18%. (d) Stress-strain curves of pristine samples with different nanopillar sizes, all exhibiting brittle fracture. (e) Stress-strain curves of 10× scratched samples with varying pillar sizes, all displaying stable plastic deformation beyond ~30% strain. (f) Stress-strain curves of 20× scratched samples with varying pillar sizes, showing a combination of plastic deformation and fracture, and an overall reduced plasticity compared to the 10× scratched case. The videos can be found in **Supplementary Materials/Videos**.

To identify the nature of the dislocations introduced in the 10× scratched KTO sample (the intermediate density in this condition merits its selection), we first employed weak-beam dark-field TEM. This technique allowed us to unambiguously confirm the presence of both edge and screw dislocations, as illustrated in **Figure S9**. Such direct observation is key for further analysis of how these distinct dislocation types contribute to the observed mechanical responses. To elucidate the atomic-scale origin of the ductile regime observed in KTO, we performed high-resolution high-angle annular dark-field (HAADF)-STEM and energy dispersive spectroscopy (EDS) analyses on the 10× scratched sample, which exhibited the peak performance in plasticity. **Figure 4a** reveals a distinct antiphase boundary (APB) within the lattice, terminating at two edge dislocations in climb dissociation configuration with ½<110> Burgers vectors (**Figure 4b**). This configuration evidences a strong interaction between dislocations and planar defects, suggesting that APBs may serve as barriers for dislocation motion [40]. Strain mapping via geometric phase analysis (GPA, **Figure 4c**) reveals pronounced local lattice distortion in the vicinity of the dislocation cores, confirming their strain-bearing nature. Notably, the APB structure (**Figure 4d**) displays an ordered atomic configuration distinct from classical dislocation lines, implying a unique core structure possibly stabilized by ionic mismatch or charge redistribution [23]. HAADF-STEM imaging combined with elemental mapping (**Figures 4e-f**) further reveals localized compositional fluctuations at the screw dislocation-APB interface, particularly in the distributions of K and Ta. Such heterogeneity can effectively modulate local bonding strength and defect energetics, facilitating dislocation nucleation and/or pinning. These results point to a cooperative mechanism in which dislocation-APB interactions mediate plastic flow, enhancing ductility at moderate defect densities.

More interestingly, comparative analysis with STO [24] reveals that KTO possesses a greater propensity for forming complex dislocation-APB networks, consistent with its enhanced capacity for dislocation multiplication (with much higher dislocation density under the same mechanical loading conditions) and higher threshold of fracture strain. This structural flexibility may arise from the more covalent Ta–O bonding and less rigid octahedral framework in KTO, allowing for easier accommodation of lattice distortions, based on the most recent molecular dynamics simulations [23]. Together, these atomic-resolution insights indicate that the BDB transition in KTO appears more than merely a function of dislocation density, but intricately governed by the structural change and chemical fluctuation of dislocation-defect interactions, providing a mechanistic basis for defect-informed design of deformable ceramics.

***Updated version.***

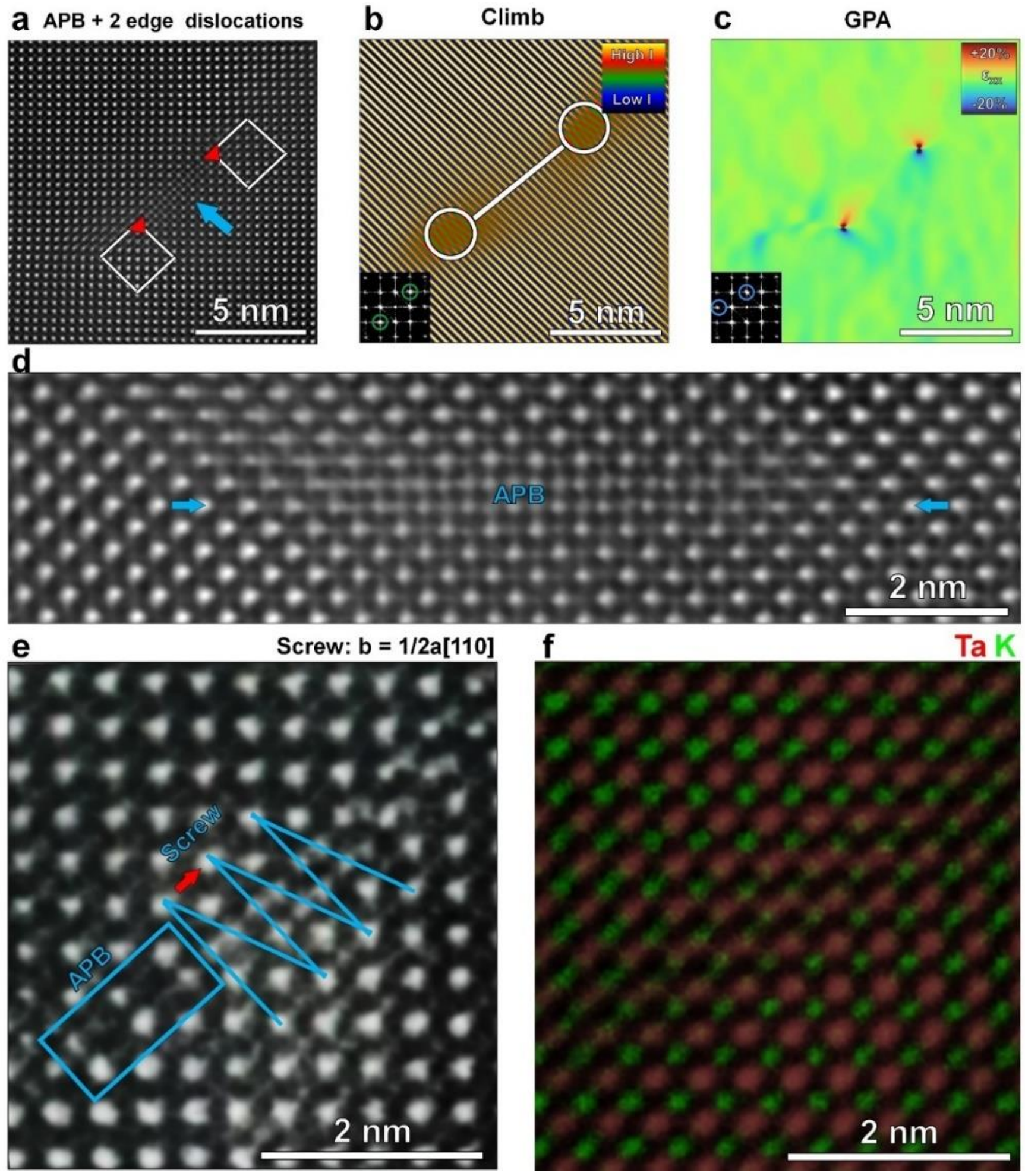

**Figure 4.** Atomic-resolution HAADF-STEM and EDS mapping of the 10× scratched KTO crystal: (a) Atomic-resolution HAADF-STEM image of KTO along the <001> zone axis, illustrating an anti-phase boundary (APB), as indicated by the blue arrow. The APB is terminated by two edge dislocations in climb dissociation configuration with Burgers vectors of 1/2<110>. (b) Inverse fast Fourier transform (IFFT) image highlighting the positions of the two edge dislocations. (c) GPA strain map corresponding to panel (a), visualizing the strain distribution associated with the edge dislocations. (d) Magnified view of the APB structure from panel (a), displaying an atomic arrangement that differs from conventional edge or screw dislocations. (e) Atomic-resolution HAADF-STEM imaging and (f) EDS mapping of both the screw dislocation and the APB region, highlighting local compositional variations.

Next to the mechanical properties, we further investigate the functional properties tuned by dislocations by showcasing the thermal conductivity (see **Materials & Methods**). **Figure 5a** compiles the yield strength, fracture strain, and thermal conductivity for KTO as a function of increasing dislocation density. This more complete picture reveals two interesting features. First, the thermal conductivity decreases monotonically as the dislocation density increases. This is expected as the large strain field around dislocations have long been proposed [41-43] and recently evidenced by room-temperature observation to effectively scatter phonons to reduce thermal conductivity [33]. Note that the slope becomes steeper with the increased dislocation density above ~$10^{13}$/m$^2$. This can be attributed to the fact that, in addition to the dislocations as line defects for phonon scattering, the increased densities in APBs as 2D defects (**Figure 4**) will add to the phonon scattering as the dislocation density increases. The contribution of boundary scattering has been extensively modelled and validated elsewhere [35, 44]. Second, the mechanical responses exhibit a clear non-monotonic change. In the samples with low dislocation densities (<$10^{13}$ m$^{-2}$), brittle fracture with high fracture strength (>7 GPa) and negligible plastic strain (< 1%) prevails. As dislocation density increases, the capability to plastically deform improves markedly, reaching a maximum at ~$10^{14}$ m$^{-2}$, where plastic strains approaching 25% without catastrophic failure. Beyond this threshold, the trend in plastic deformability starts to reverse, with a fracture strain declining sharply.

For the sake of generality, in **Figure 5b** we present the data for STO, with an analogous behavior observed. This demonstrates that such dislocation density-dependent competition between the mechanical and functional properties is not unique to KTO, but should hold true for other perovskite oxides provided that dislocations can be engineered. Nevertheless, KTO exhibits a more dramatic change especially in the fracture strain compared to STO, due to the fact that a one order of magnitude higher dislocation density is achievable in KTO.

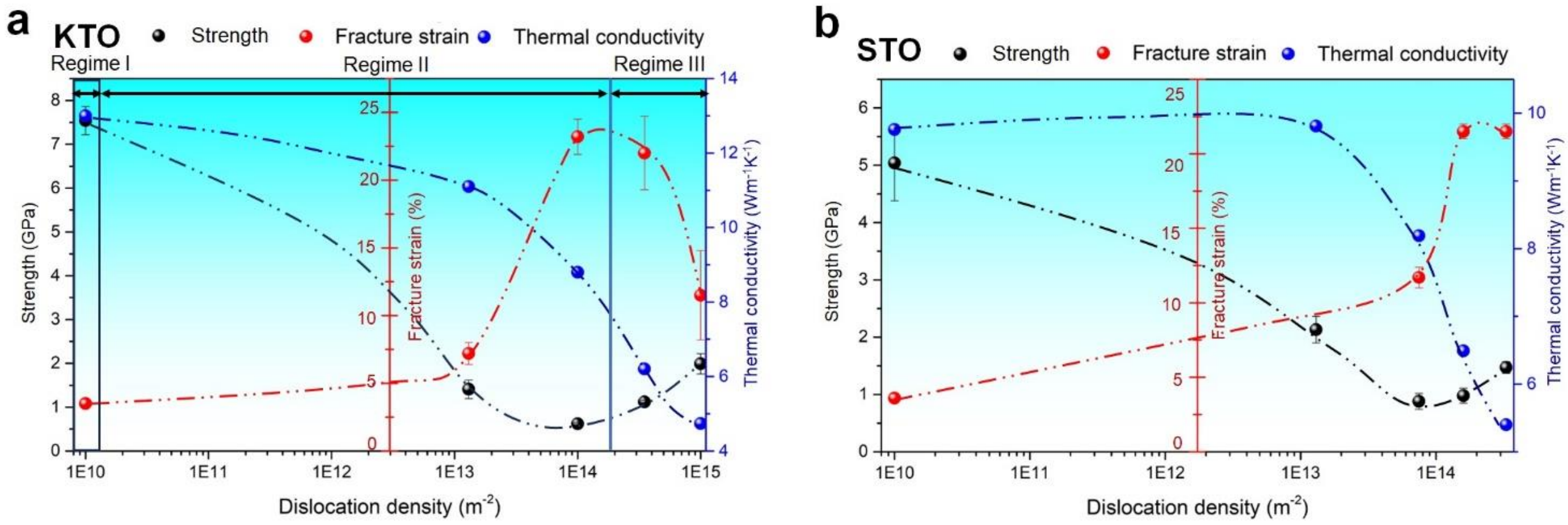

**Figure 5.** (a) Plot showing the variation of the yield strength and fracture strain in micropillar compression, as well as thermal conductivity as a function of dislocation density for KTO crystal. It shows that thermal conductivity decreases monotonically with increasing dislocation density, while mechanical properties exhibit a non-monotonic evolution, i.e., transitioning from brittle fracture at low densities to enhanced plasticity at intermediate levels, before degrading again at high densities. Note the three different regimes I, II, III for indications. (b) Similar plot as in (a) for STO crystal. With mechanical data reproduced from recent literature [24], and the thermal conductivity measured for this work.

***Updated version.***

## Discussions

### *BDB transition*

The underlying mechanism for the V-shaped curve for the strength as a function of dislocation in STO [11] has been interpreted within the framework of dislocation mechanics similar to metals by El-Awady et al. [27]. Here, we focus on the BDB transition in KTO at room temperature, where the brittle fracture at both ends of the curve at both low (~$10^{12}$ $m^{-2}$) and extremely high (~$10^{15}$ $m^{-2}$) dislocation densities features a distinction in ceramics compared to metals.

In metals, particularly BCC metals such as tungsten, the brittle-ductile transition is controlled by dislocation mobility as function of temperature [45]. This is not the case in ceramics as their plasticity is restricted rather by dislocation nucleation [11, 46]. This shift of the bottleneck is caused by the fundamentally different processing routes for forging metals and sintering ceramics. Conventionally sintered ceramics by high-temperature firing or grown crystals exhibit approximately 4-5 orders of magnitude lower dislocation densities compared to metals. Down to the nano-/microscale, there is barely one dislocation existing in the deformation volume in ceramics (Regime I, **Figure 5a**). Hence, for plasticity to occur it requires dislocation nucleation at a stress level approaching a fraction of the shear modulus to break the bonds in shear deformation [47, 48]. This picture has been fundamentally changed by *dislocation seeding* to circumvent the high nucleation barrier and shift the restricting process to dislocation motion and multiplication, which proves much easier in dozens of ceramics [10, 11, 14]. In particular, dislocation motion and multiplication are operating more effectively to carry the plasticity in the intermediate range (Regime II, **Figure 5a**) for both representative perovskite oxides KTO and STO (~$10^{12}$-$10^{14}$ $m^{-2}$). However, at ultrahigh densities (Regime III, **Figure 5a**), dislocation-dislocation interactions and dislocation forest hardening, as well as entanglement with APBs in the case of KTO, start to dominate. The much-limited available slip systems in perovskite oxides at room temperature (6 physically distinct but only 2 independent slip systems in STO and KTO with the cubic structure [1]) and limited ability to cross slip due to the dissociation of dislocations at ambient conditions lead to more severe dislocation jamming to nucleate cracks [24, 49-52]. In short, the dynamics between available mobile dislocations and immobile pinning structures (being either sessile dislocations or APBs) likely governs the observed BDB transition in KTO.

***Updated version.***

### *Size effect*

Both tests at nanoscale and microscale, being *in situ* or *ex situ*, validates that the density-mediated properties are consistent at small scales. Since both KTO and STO are experimentally demonstrated to be capable of room-temperature bulk plasticity [20-22], and the dislocation seeding approach using Brinell indentation and scratching to generate meso/macroscale plastic zone size applies [21, 53], it is in principle feasible to scale it up to meso/macroscale for the density-dependent measurement. Nevertheless, consider the high cost and limited availability of large single crystals of such perovskite oxides, one of the most promising application scenarios shall be at the small scale for potential MEMS/NEMS devices using dislocations, matching the length scales investigated here.

### *Mechano-functional tradeoff*

The finding of the dislocation density-dependent BDB transition deserves attention for the endeavors in dislocation-tuned functionality of ceramics, for which ultrahigh dislocation density is desirable. Here we mainly demonstrate the thermal conductivity as a monotonic function of dislocation density over the whole achievable range, for both STO and KTO. Nevertheless, dislocation-tuned electrical conductivity has been reported to exhibit analogous monotonic behavior [12, 18, 31, 54]. Given the scenarios, next to the superior functional performance, where excellent mechanical integrity and structural robustness of the components is critical, the tradeoff between the mechanical and functional properties, namely, *mechano-functional tradeoff* mediated by dislocations in ceramics shall be borne in mind. This is relevant for designing emerging functional dislocation-based devices operating in multifield-coupled and particularly mechanical load-bearing environments.

## Conclusions

This work provides a new mechanistic framework for dislocation engineering in functional oxides. For the first time, the direct comparison between dislocation-tuned mechanical and functional properties triggered a reassessment of the role of dislocations in ceramics. By demonstrating a controllable dislocation-density-based BDB transition in perovskite oxides, we establish that ductility in functional oxides is not a binary state but rather a dynamic outcome governed by defect topology, density, and interaction mechanisms. This BDB transition, distinct from metallic systems and any other ductile ceramics reported before, reveals a fundamental limit in dislocation-mediated toughening: while defect engineering can enhance ductility, there exists a critical density beyond which the crystal loses its ability to accommodate strain plastically. Understanding this tipping point is essential for optimizing mechanical robustness in oxide-based devices, particularly those involving mechanical/multi-field loading. Our findings advance the defect-mechanics-functionality paradigm beyond traditional endeavor in ceramics and open new avenues for engineering functional oxides.

## Supplementary Materials

**This part contains:**

**A) Materials & Methods**

**B) 10 supplementary figures (Figure. S1-S10)**

**C) 12 supplementary videos (Video1-12)**

***Updated version.***

## A) Materials & Methods

**Materials and Methods:**

***Mechanical seeding of dislocations***

The (001)-oriented $KTaO_3$ single crystals were acquired from Hefei Ruijing Optoelectronics Technology Co., Ltd. (Anhui, China). The $KTaO_3$ thin plates with a dimension of 5 mm × 5 mm × 1 mm were single-side polished to a surface roughness lower than 2 nm. A friction and wear machine (RTec, MFT) equipped with a 3 mm diameter $\alpha$-$Al_2O_3$ ball was employed to generate different wear tracks by tuning the cyclic scratching passes (**Figure S10**). The scratching load was set to 8 N, with a scratching speed of 0.2 mm/s and distances of 2 mm. The optimized parameters ensure no surface cracks were generated.

***TEM analysis***

The TEM specimens were lifted out along the scratching direction in the center of the wear tracks using a dual-beam focused ion beam (FIB) in an SEM (Helios Nanolab 600i, FEI, Hillsboro, USA). ABF-STEM images were captured using a TEM instrument (FEI Talos F200X G2, Thermo Fisher Scientific, USA) operating at 200 kV. A probe semi-convergence angle of 10.5 mrad and inner and outer semi-collection angles of 12-20 mrad were used. For atomic-scale imaging, HAADF-STEM imaging was then performed in a double aberration-corrected transmission electron microscope (Titan Themis G2, FEI, Netherlands) operated at 300 kV. A probe semi-convergence angle of 17 mrad was utilized, with inner and outer semi-collection angles ranging from 38 to 200 mrad.

***Pillar compression***

For micropillar compression tests, all the micropillars were prepared in the FIB with decreased ionic current of 0.79 nA-0.43 nA-0.23 nA-80 pA and a voltage of 30 kV. The *ex situ* micropillar compression tests (pillars with diameters of 1 and 3 μm) were performed on the nanoindenter equipped with a ~5 μm diameter flat punch indenter (Hysitron TI Premier, Bruker, USA). This excludes the potential electron beam effect on the mechanical properties. The surface morphologies of the deformed micropillars were imaged by the scanning electron microscope (SEM, Apreo2 S Lovac, USA). The *in situ* nanopillar compression experiments were performed on the TEM (Thermo

***Updated version.***

Fisher, Talos F200X G2, USA), equipped with a PI95 flat punch picoindenter (~1 μm in diameter). All the compression modes were set to displacement-controlled, with a strain rate of ~$1 \times 10^{-3}$ $s^{-1}$.

***Thermal conductivity measurement***

The room temperature thermal conductivities of the $SrTiO_3$ and $KTaO_3$ single crystals with mechanically seeded dislocations were measured using the time-domain thermoreflectance (TDTR) method [55]. Before the measurements, the mechanically seeded $SrTiO_3$ and $KTaO_3$ thin plates were coated with an aluminum (Al) layer with a thickness of ~70 nm. All the thermal conductivity data of the mechanically seeded dislocation samples were collected in the middle of the wear tracks. Detailed data processing can be found in Ref. [33].

**B) 10 supplementary figures (Figure. S1-S10)**

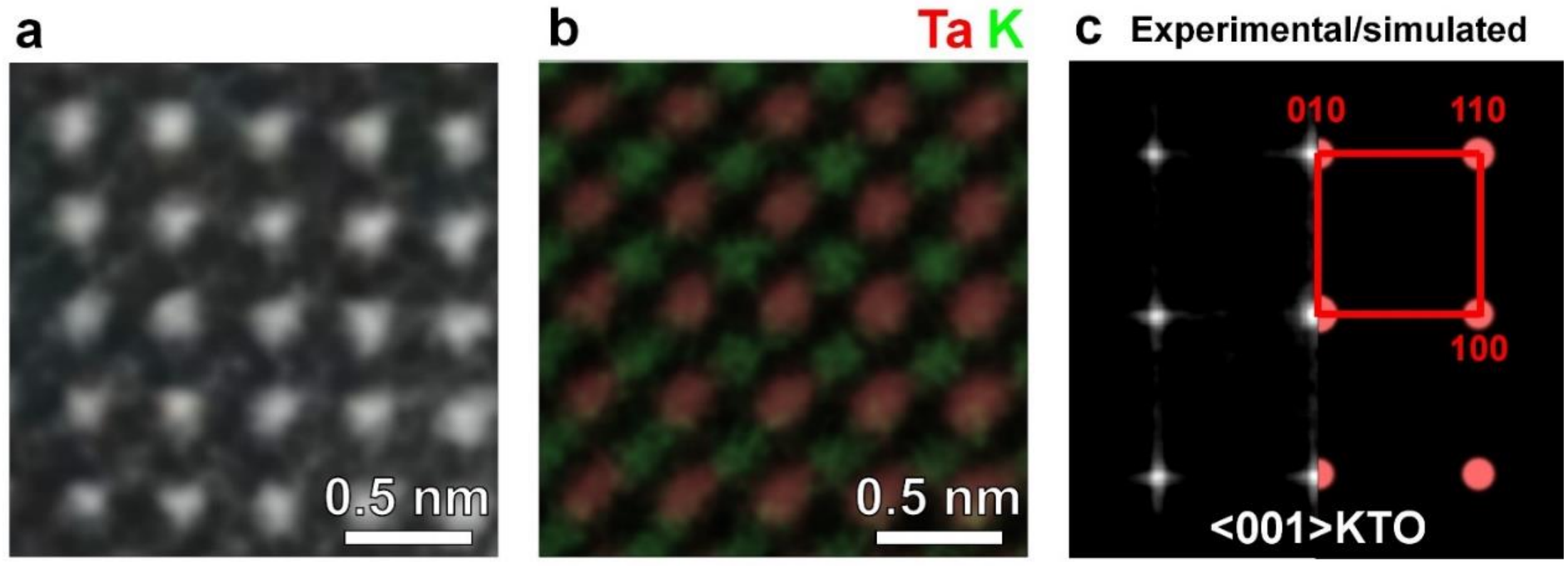

**Figure S1.** Atomic-scale characterization of the KTO reference sample: (a) A HAADF-STEM image at atomic resolution. (b) Corresponding EDS elemental mapping of potassium (K) and tantalum (Ta). (c) Electron diffraction pattern with simulated data of KTO along the <001> zone axis.

***Updated version.***

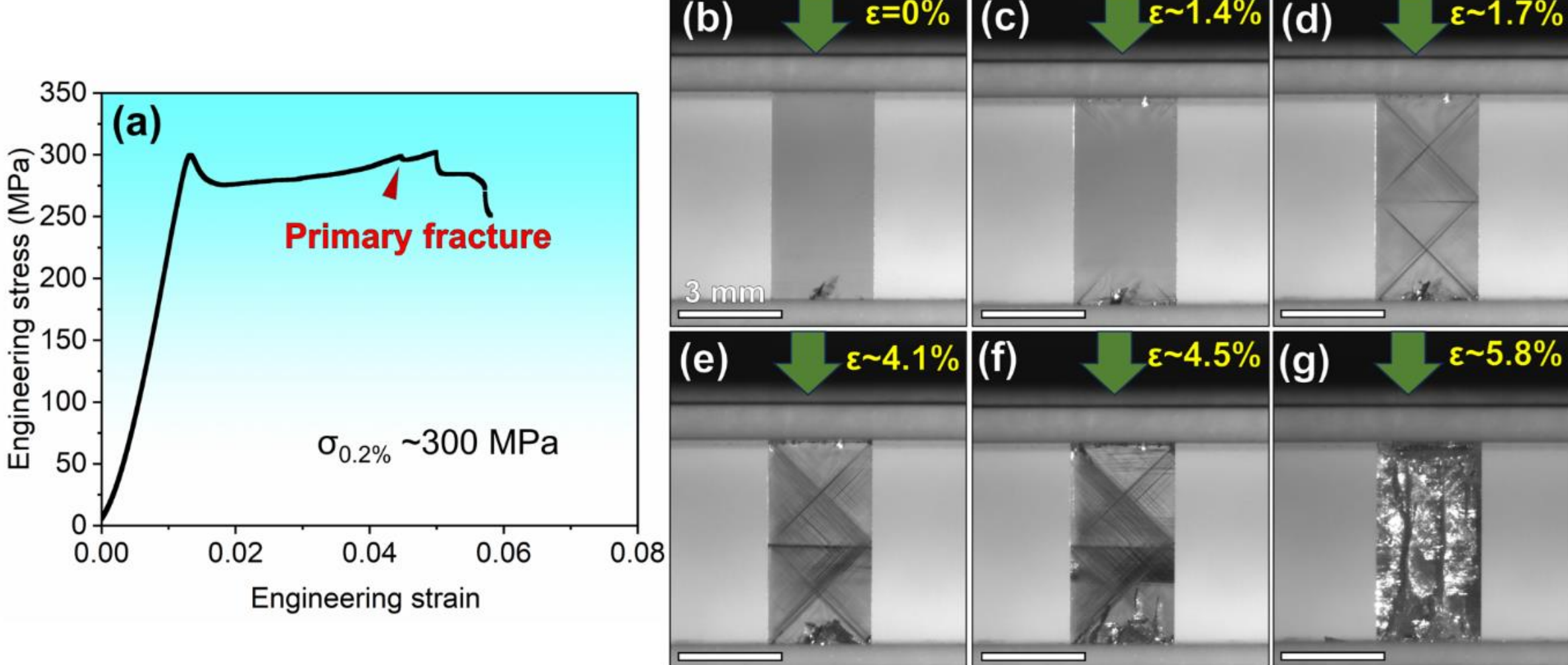

**Figure S2.** Room-temperature bulk compression of single-crystal KTO along the <001> loading direction with a strain rate of ~$10^{-4}$/s. The slip traces (45-degree tilt dark lines with respect to the loading axis) indicate the <110> slip directions on the {110} slip planes. The sample is capable of ~5% plastic strain without critical fracture, and the lowest yielding point is around 270 MPa.

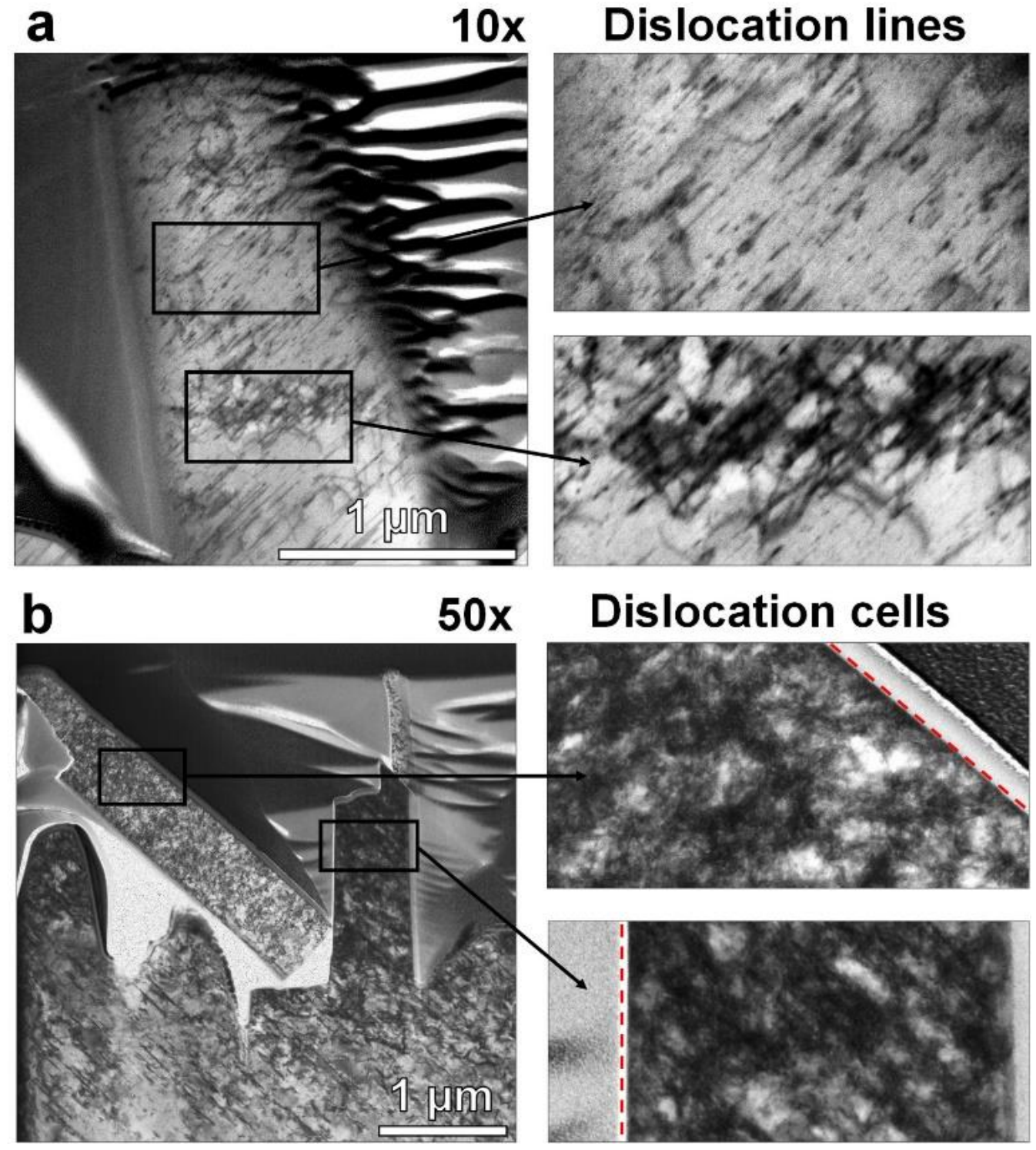

**Figure S3.** Structural characterization of micropillars with two typical dislocation densities after compression: (a) An ABF-STEM image showing the morphology of the 10× scratched sample after compressive deformation up to 30% strain. (b) An ABF-STEM image displaying the fractured morphology of the 50× scratched sample after compressive deformation to 4.2% strain.

***Updated version.***

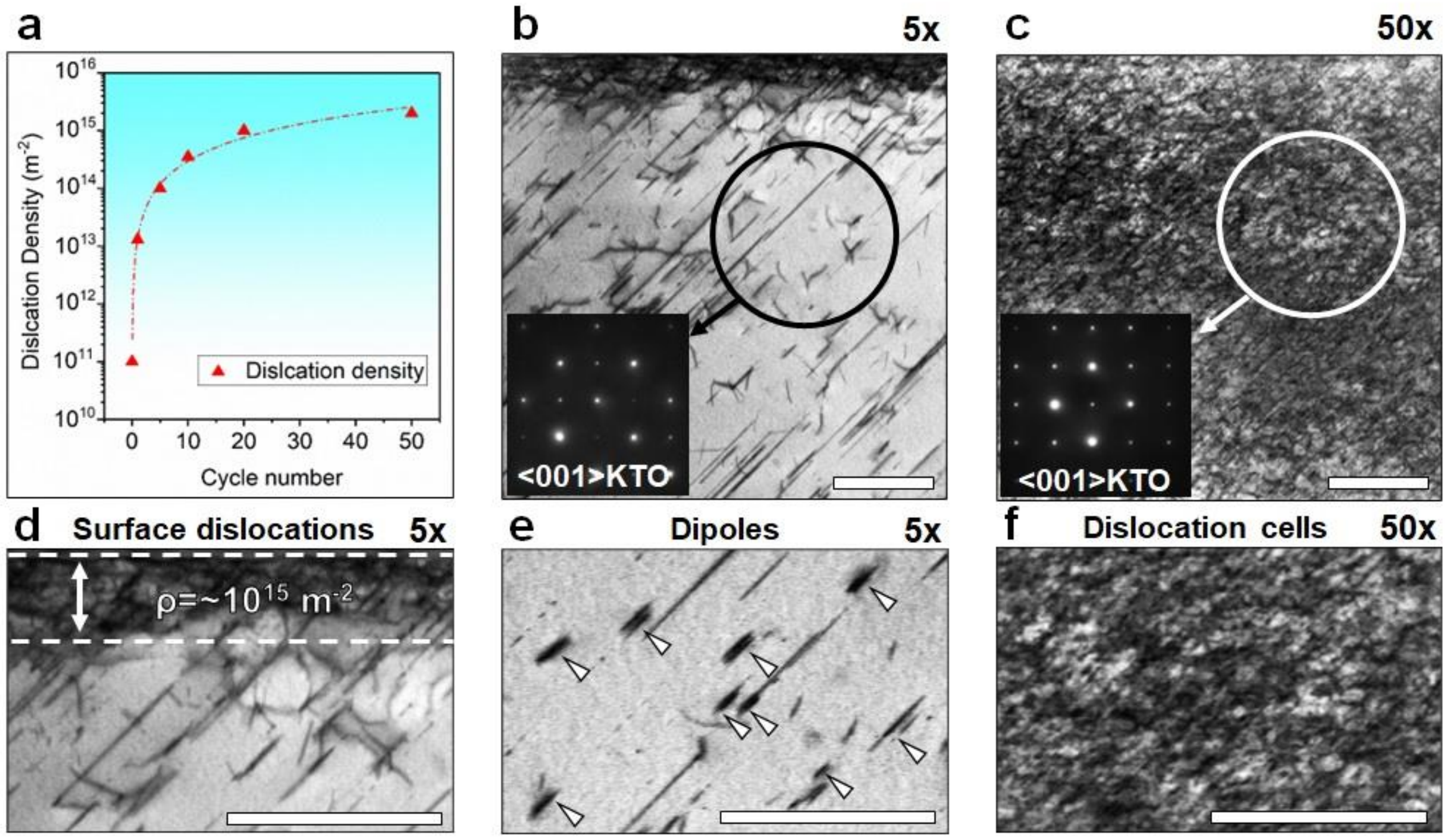

**Figure S4.** (a) Relationship between dislocation density and the number of scratching cycles. (b) Low-magnification image of the 5× scratched sample. (c) Low-magnification image of the 50× scratched sample. (d) Magnified view of the 5× sample showing a gradient distribution of surface dislocations. (e) High-resolution image of the 5× sample revealing the presence of stacking faults or dislocation dipoles. (f) High-resolution image of the 5× sample illustrating the formation of dislocation cell structures at high dislocation density.

***Updated version.***

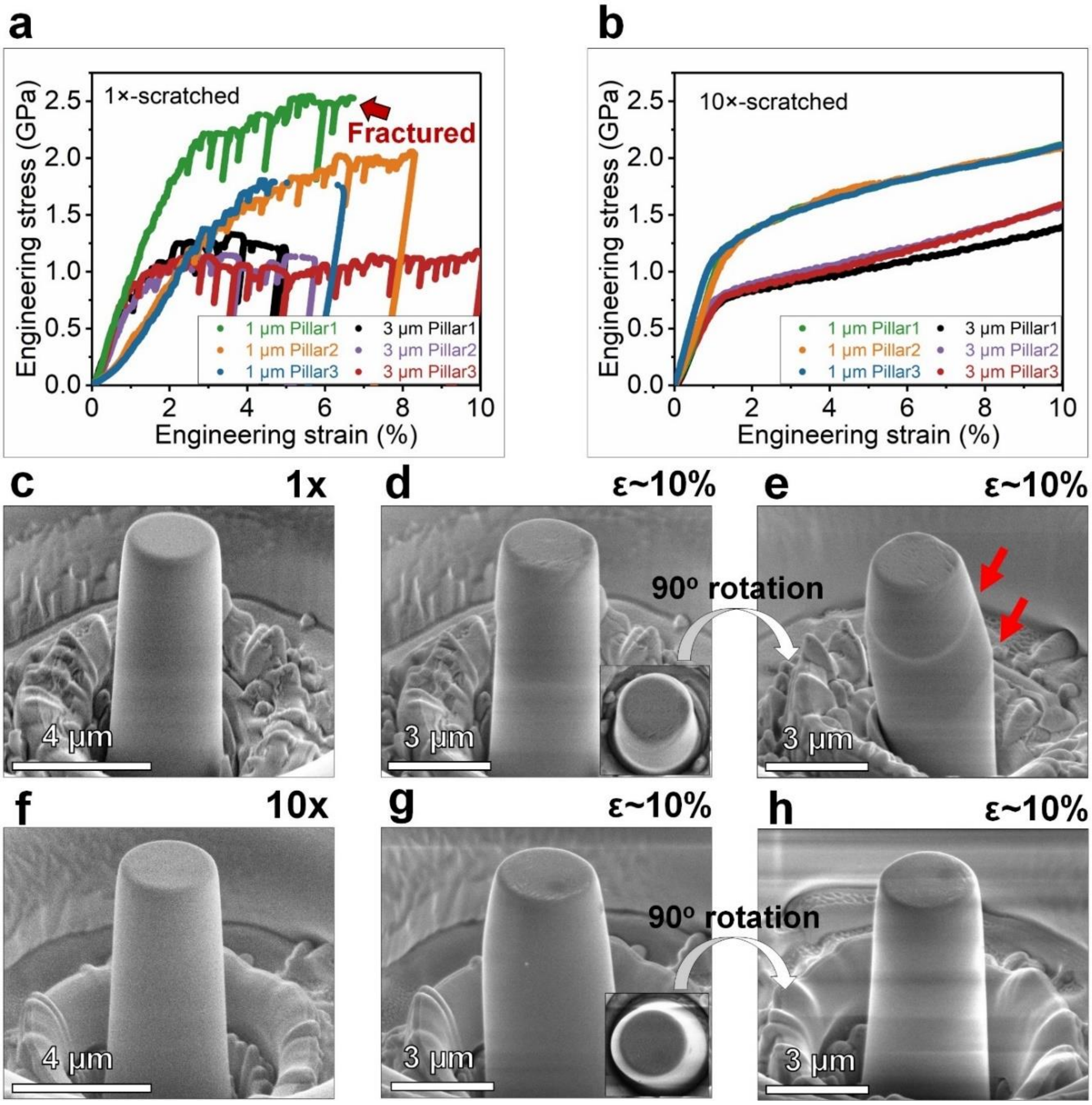

**Figure S5.** Size effect mechanics and structural characterization of micropillar compression: (a) Engineering stress-strain curves of 1× scratched samples under size effect (pillar diameters: 1-3 μm). (b) Engineering stress-strain curves of 10× scratched samples under size effect (pillar diameters: 1-3 μm). (c) A SEM image of an undeformed 3 μm micropillar from the 1× sample. (d) A SEM image of the same pillar after 10% compressive strain. The inset is the top view of the pillar. (e) Side-view SEM image of the same pillar rotated by 90°, with red arrows indicating visible slip bands. (f) SEM image of an undeformed 3 μm micropillar from the 10× sample. (g) SEM image of the 10× sample's pillar after 10% compressive strain. The inset is the top view of the pillar. (h) Side-view SEM image of the deformed pillar after 90° rotation.

***Updated version.***

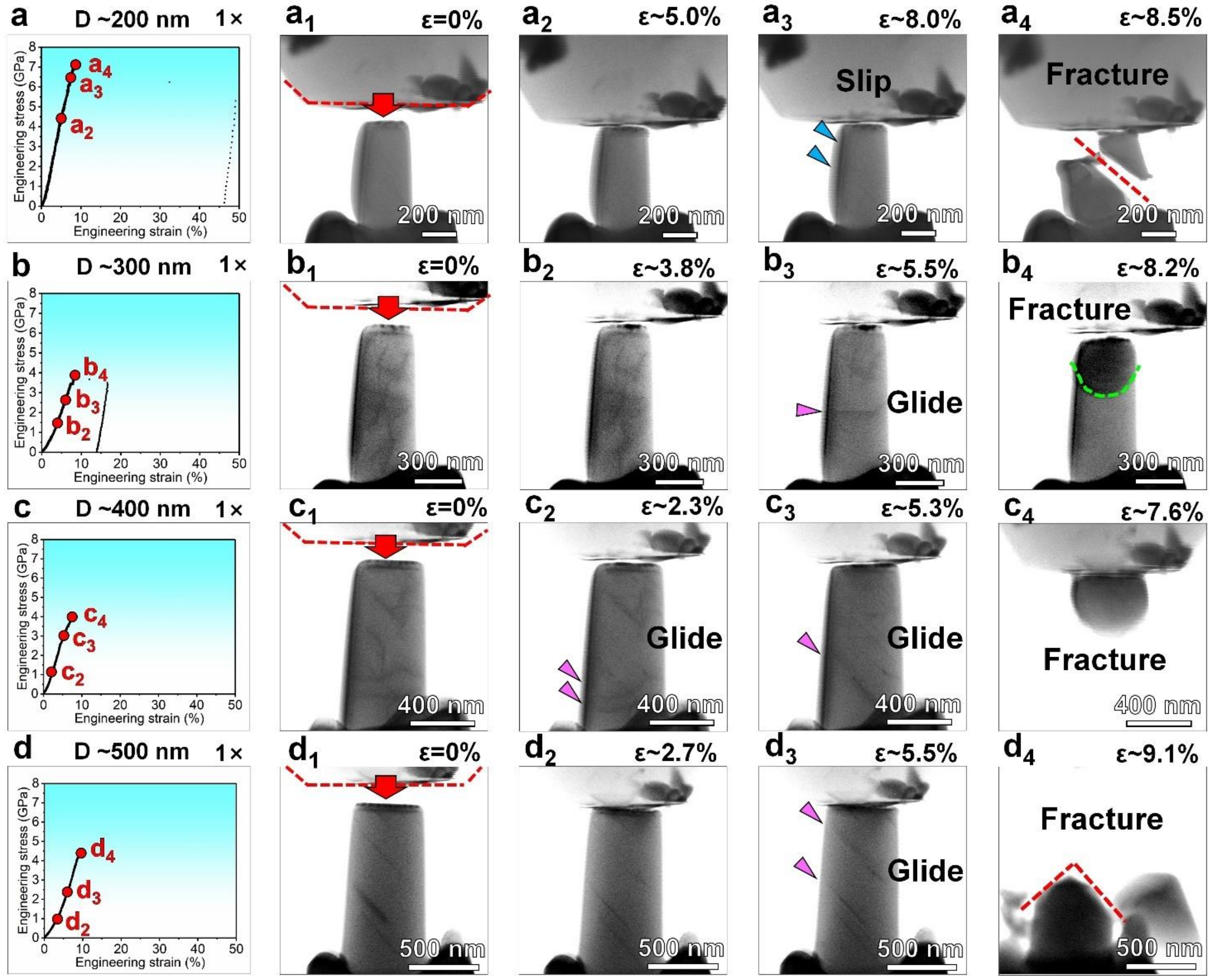

**Figure S6.** Engineering stress-strain responses and in-situ deformation behavior of 1× nanopillars under compression observed via ABF-STEM imaging: (a) Compressive stress–strain profile of the 1× sample with a diameter of 200 nm. (a1–a4) Sequential ABF-STEM images capturing the real-time deformation process of the 200 nm pillar. Fracture occurs at a maximum strain of approximately 8.5%. (b) Mechanical response of the 300 nm 1× nanopillar under uniaxial compression. (b1–b4) In-situ ABF-STEM frames illustrating the deformation evolution. Dislocation glide (highlighted by the purple arrow) is clearly evident, with fracture taking place at around 8.2% strain. (c) Stress-strain behavior of the 400 nm 1× sample during compression. (c1–c4) Time-resolved ABF-STEM images revealing dislocation glide (purple arrow) and subsequent fracture at approximately 7.6% strain. (d) Load–displacement curve converted to engineering stress-strain for the 1× pillar with a 500 nm diameter. (d1–d4) ABF-STEM snapshots tracking the deformation sequence. Dislocation motion (indicated by purple arrows) is distinctly observed, and fracture occurs at a strain level of about 9.1%.

***Updated version.***

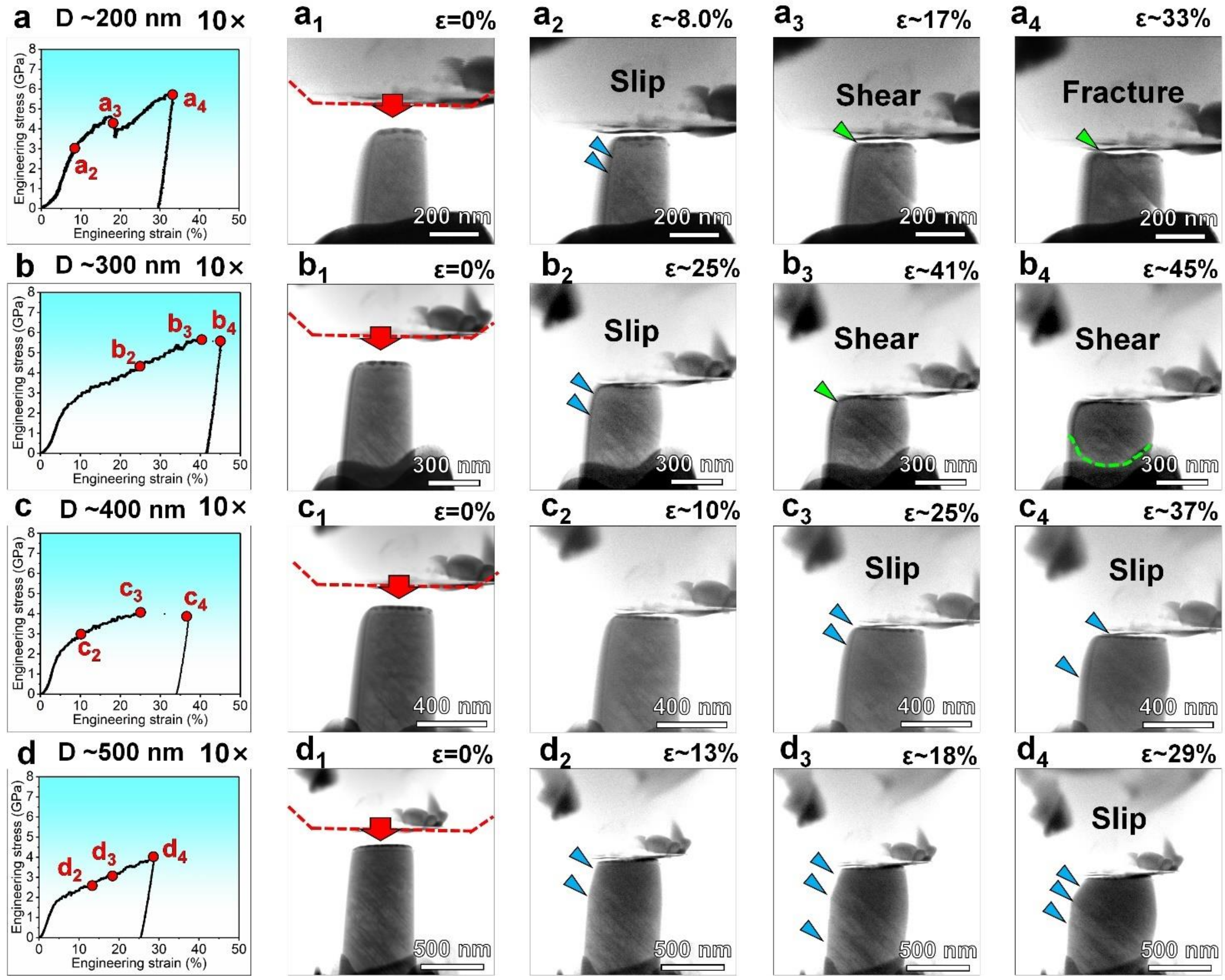

**Figure S7.** Mechanical performance and in-situ deformation characteristics of 10× nanopillars under compression, captured via ABF-STEM (dislocation slip indicated by blue arrows and shear marked by green arrows): (a) Stress-strain curve of the 10× sample with a 200 nm diameter under compressive loading. (a1–a4) ABF-STEM image sequence documenting the deformation progression of the 200 nm pillar. Fracture occurs at an ultimate strain of approximately 33%. (b) Compressive behavior of the 300 nm 10× nanopillar. (b1–b4) In-situ ABF-STEM frames detailing the deformation path, where both dislocation slip and shear activity are prominently observed. Shearing takes place at around 45% strain. (c) Mechanical response of the 400 nm sample during compression. (c1–c4) Time-lapse ABF-STEM images capturing the onset and development of dislocation glide (blue arrow), followed by plastic deformation without failure at roughly 37% strain. (d) Converted stress–strain profile from load-displacement data for the 500 nm pillar. (d1–d4) ABF-STEM views revealing the dynamic deformation sequence, where intense dislocation slip (highlighted by blue arrows) precedes plastic deformation at a strain of approximately 29%.

***Updated version.***

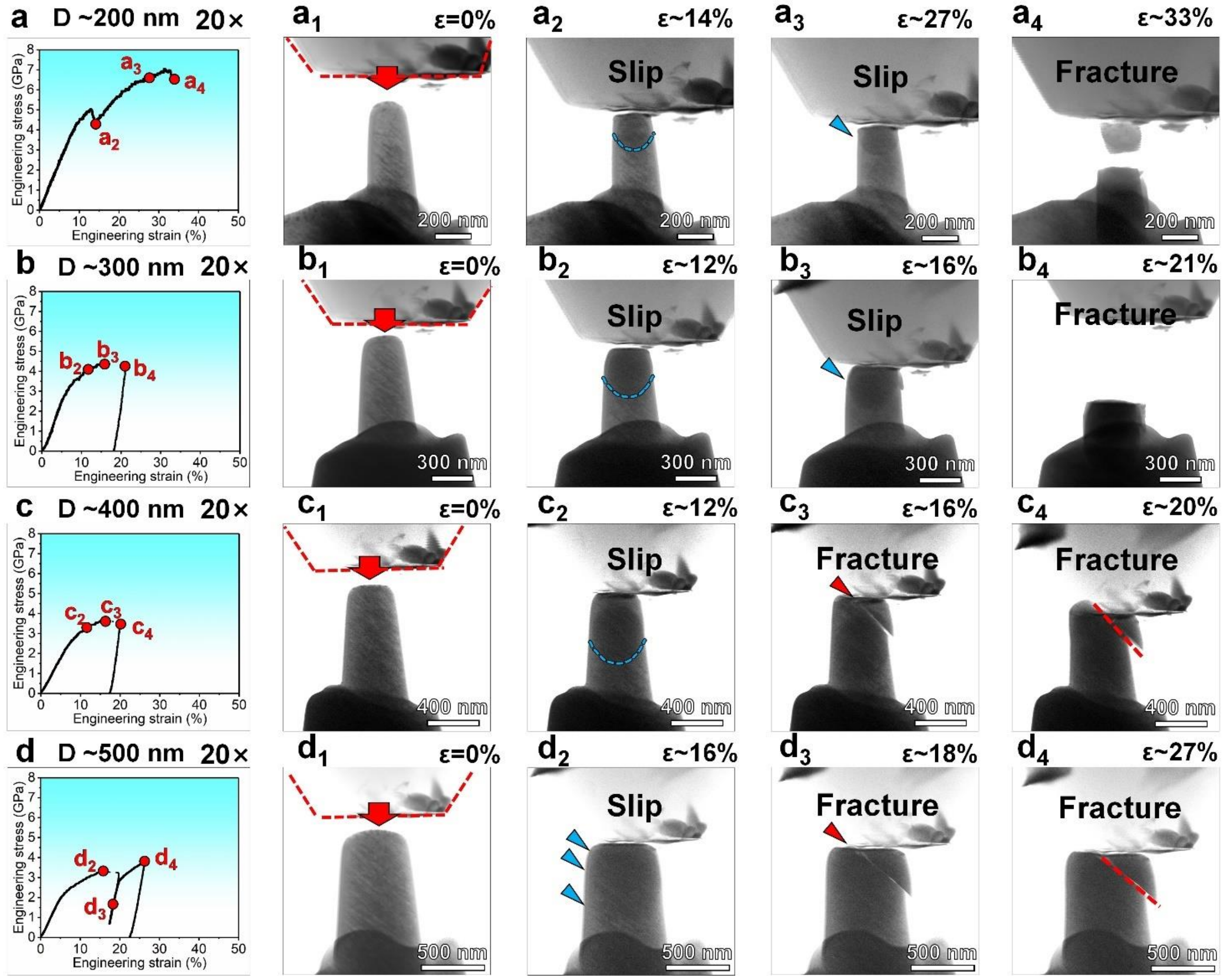

**Figure S8.** Mechanical response and in-situ deformation behavior of 20× nanopillars under compression, observed via ABF-STEM (dislocation slip denoted by blue arrows and fracture by red arrows): (a) Compressive stress-strain profile of the 20× sample with a 200 nm diameter. (a1–a4) ABF-STEM series illustrating the progressive deformation of the 200 nm pillar, which eventually fractures at a strain of ~33%. (b) Deformation characteristics of the 300 nm nanopillar under axial loading. (b1–b4) Real-time ABF-STEM snapshots showing clear evidence of dislocation slip and shear activity, with fracture occurring around 21% strain. (c) Compressive performance of the 400 nm pillar. (c1–c4) Sequential STEM images capturing the initiation and propagation of dislocation motion (blue arrow), culminating in failure at approximately 20% strain. (d) Stress-strain relationship derived from load-displacement measurements for the 500 nm sample. (d1–d4) STEM visualizations mapping the deformation evolution, where significant dislocation slip (marked in blue) precedes fracture at a strain close to 27%.

***Updated version.***

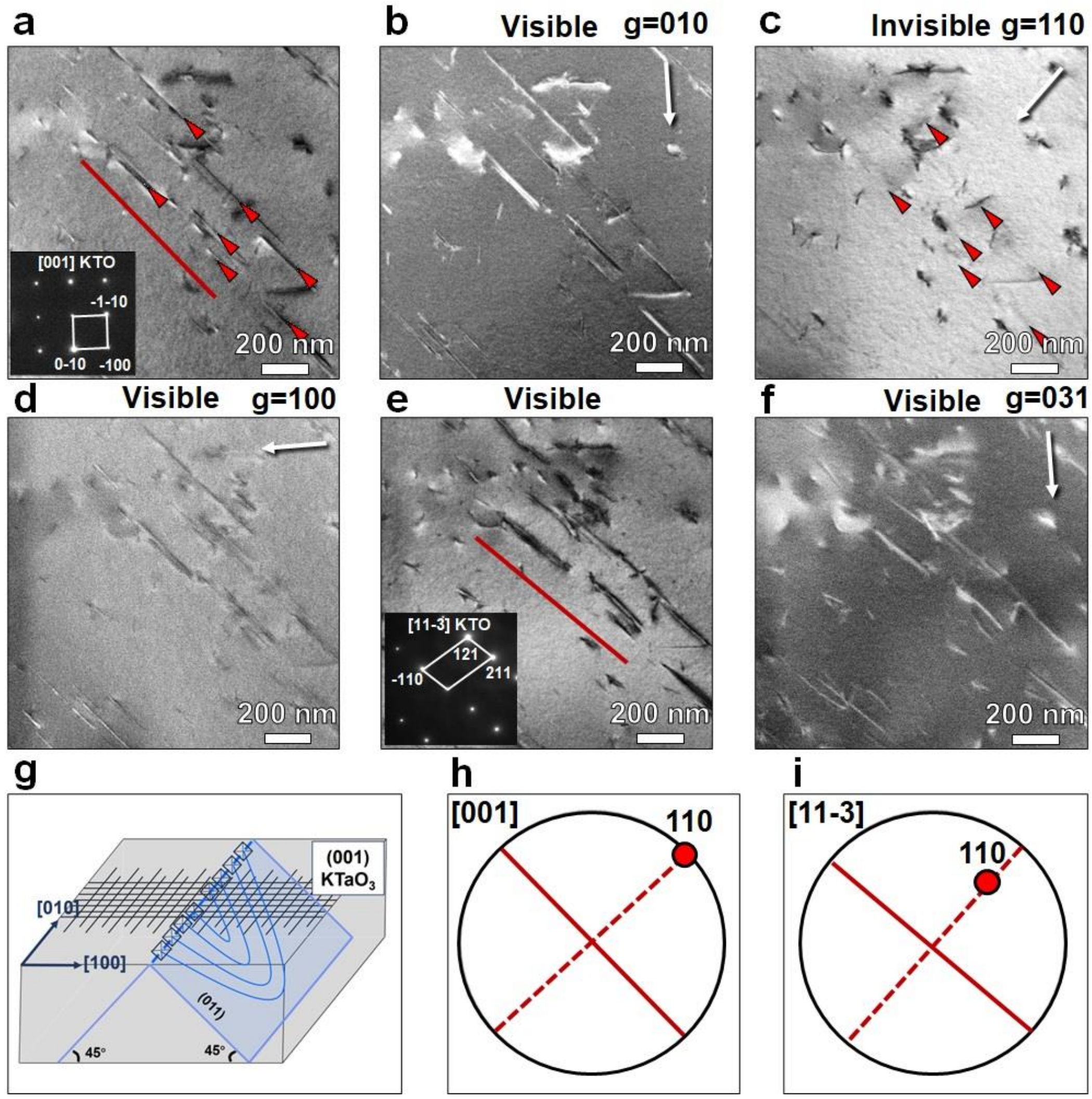

**Figure S9.** Determination of the Burgers vector and slip system in the 10× scratched KTO sample: (a) Dislocation structure of single-crystal KTO along the <001> zone axis. The red arrow indicates a dislocation line oriented at 45°, with the slip plane identified as {110}. (b) Under a diffraction vector ***g*** = 010, the dislocation line remains visible (no extinction). (c) Under ***g*** = 110, the dislocation line is extinguished. (d) Under ***g*** = -1-10, the dislocation line remains visible (no extinction). (e) Dislocation structure of single-crystal KTO along the <11-3> zone axis. (f) Under ***g*** = 031, the dislocation line remains visible (no extinction). (g) Schematic illustration of the activated slip systems in single-crystal KTO. (h) Pole figure of single-crystal KTO along the <001> axis, where red poles indicate edge-on slip planes. (i) Pole figure of single-crystal KTO along the <11-3> axis, where red poles represent inclined slip planes.

***Updated version.***

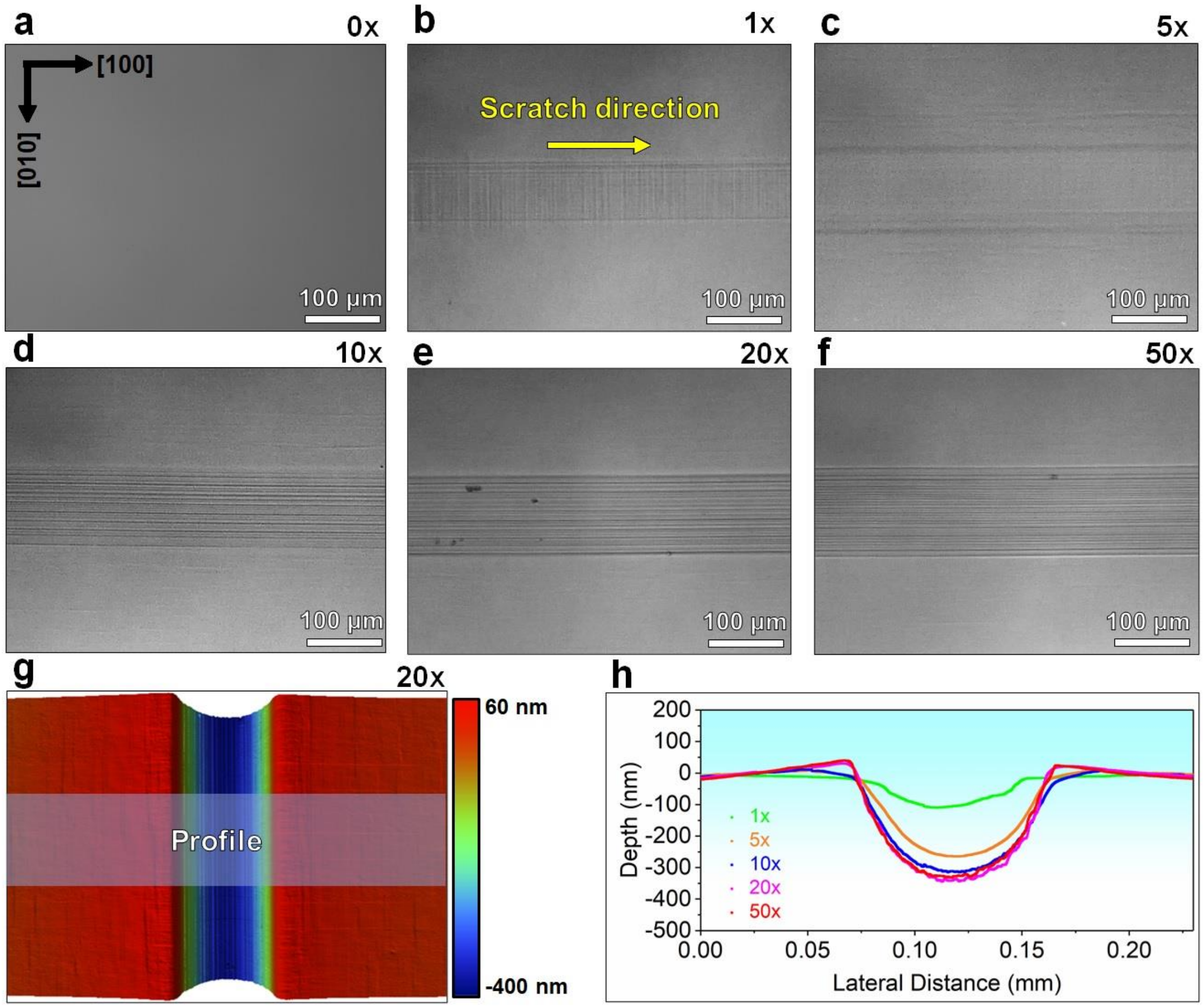

**Figure S10.** Optical microscopy images of KTO samples subjected to different numbers of scratching cycles: (a) 0×; (b) 1×; (c) 5×; (d) 10×; (e) 20×; and (f) 50×. (g) Surface topography of the 20× scratched sample obtained by white-light interferometry. (h) Depth profiles of surface indentations along the Z-axis for samples with varying scratch cycles.

### C) 12 supplementary videos (Video1-12)

Please see uploaded videos: Videos 1-4 for low dislocation density (1× scratching) for 4 different pillar diameters, and Videos 5-8 for intermediate dislocation density (10× scratching) for 4 different pillar diameters, and Videos 9-12 for high dislocation density (20× scratching) for 4 different pillar diameters.

### Acknowledgement:

X. Fang acknowledges the funding for fundamental research from the European Research Council (ERC Starting Grant, Project MECERDIS, grant No. 101076167). Views and opinions expressed are, however, those of the authors only and do not necessarily reflect those of the European Union or the European Research Council (ERC). Neither the European Union nor the granting authority can be held responsible for them. W. Lu acknowledges the support by Shenzhen Science and Technology Program (grant no JCYJ20230807093416034), the Open Fund of the Microscopy Science and Technology-Songshan Lake Science City (grant No. 202401204), National Natural Science Foundation of China (grant No. 52371110) and Guangdong Basic and Applied Basic Research Foundation (grant No. 2023A1515011510 and 2024B1515120036). We thank Prof. Bo Sun and his team at Tsinghua SIGS and Guangdong Provincial Key Laboratory of Thermal Management Engineering & Materials for the thermal conductivity measurement, and Jinxue Ding for analyzing the density-dependent thermal conductivity data.

**Author contribution:**

X. Fang and W. Lu conceptualized the idea. W. Lu and X. Fang designed and supervised the project. J. Zhang performed experiments, collected and analyzed the mechanical data. All authors discussed the results, wrote the manuscript, and approved for submission.

**Conflict of interest:** The authors declare no conflict of interest.

**Data availability:** All data related to this work has been included in the main text or the supplementary materials.

*Updated version.*